\documentclass{aa}
\usepackage[latin1]{inputenc}
\usepackage{natbib}
\usepackage{theorem,algorithm,algorithmic}
\usepackage{amssymb,amsmath}
\usepackage{psfig,epsfig,amssymb,alltt}
\usepackage{latexsym}
\usepackage{graphicx}
\graphicspath{{PS/}}
\usepackage{psfig}
\psfigurepath{./PS:.}
\usepackage[english]{babel}
\usepackage{indentfirst}
\psfigurepath{./PS}
\vspace{1cm}

\newcommand{\PS}[1]{{\bf #1}}

\begin{document}
\title{Cosmological models discrimination\\ with Weak Lensing}

\author{Sandrine Pires\inst{1} \and Jean-Luc Starck\inst{1} \and Adam Amara\inst{1,2} \and \\ 
Alexandre R\'efr\'egier\inst{1} \and Romain Teyssier\inst{1}}
\institute{Laboratoire AIM, CEA/DSM-CNRS-Universite Paris Diderot, IRFU/SEDI-SAP, Service d'Astrophysique, CEA Saclay, Orme des Merisiers, 91191 Gif-sur-Yvette, France
\and
Department of Physics, ETH Z\"urich, Wolfgang-Pauli-Strasse 16, CH-8093 Z\"urich, Switzerland
}

\offprints{sandrine.pires@cea.fr}

\date{\today}

 

\abstract{Weak gravitational lensing provides a unique method to map directly the dark matter in the Universe. 
The majority of lensing analyses uses the two-point statistics of the cosmic shear field to constrain the cosmological model yielding degeneracies, such as that between $\sigma_8$ and $\Omega_m$, respectively the r.m.s. of the mass fluctuations at a scale of $8 Mpc/h$ and the matter density parameter both at $z=0$.  However, the two-point statistics only measure the Gaussian properties of the field and the weak lensing field is non-Gaussian. It has been shown that the estimation of non-Gaussian statistics on weak lensing data can improve the constraints on cosmological parameters. 
In this paper,  we systematically compare a wide range of non-Gaussian estimators in order to determine which one provides tighter constraints on the cosmological parameters.
These statistical methods include skewness, kurtosis and the Higher Criticism test in several sparse representations such as wavelet and curvelet; as well as the bispectrum, peak counting and a new introduced statistic called Wavelet Peak Counting ($WPC$). Comparisons based on sparse representations show that the wavelet transform is the most sensitive to non-Gaussian cosmological structures. It appears also that the best statistic for non-Gaussian characterization in weak lensing mass maps is the $WPC$. Finally, we show that the  $\sigma_8$-$\Omega_m$ degeneracy could be even better broken if the WPC estimation is performed on weak lensing mass maps filtered by the wavelet method, MRLens.}

\maketitle 
\markboth{Cosmological model discrimination from Weak Lensing Data}{}


\section{Introduction}

\indent Images distortion measurements of background galaxies caused by large-scale structures provides a direct way to study the statistical properties of the growth of structures in the Universe. Weak gravitational lensing measures the mass and can thus be directly compared to theoretical models of structure formation. 
Most lensing studies use the two-point statistics of the cosmic shear field because of its potential to constrain the power spectrum of density fluctuations in the late Universe \citep[e.g.][]{twopoint:maoli01,twopoint:refregier02,twopoint:bacon03,twopoint:massey05,twopoint:dahle06}.
Two-point statistics measure the Gaussian properties of the field. This is a limited amount of information since it is well known that the low redshift Universe is highly non-Gaussian on small scales. Indeed, gravitational clustering is a non linear process and in particular at small scales the mass distribution is highly non-Gaussian. Consequently, using only two-point statistics to set constraints on the cosmological model is limited. Better constraints can be obtained 
using 3D weak lensing maps \citep{threepoint:bernardeau97,threepoint:pen03,wlens:massey07}. 
An alternative procedure is to consider higher-order statistics of the weak lensing shear field enabling a characterization of the non-Gaussian nature of the signal \citep[see e.g.][]{threepoint:takada03,threepoint:jarvis04,threepoint:kilbinger05,peak:hamana04,stat:donoho04}.


In this paper, we systematically compare a range of non-Gaussian statistics. For this purpose, we focus on the degeneracy between $\sigma_8$ and $\Omega_m$,  respectively the amplitude of the matter power spectrum and the matter density parameter both at $z=0$. We investigate which statistical method is the most efficient to break this degeneracy that exists when only the two-point correlation is considered. 
A wide range of statistical methods are systematically applied on a set of simulated data
in order to characterize the non-Gaussianity present in the mass maps due to the growth of structures.
Their performance to discriminate between different possible cosmological models are compared.
For the CMB, it has  been proposed to use statistics such wavelet kurtosis or wavelet Higher Criticism to detect clusters and curvelet kurtosis or curvelet Higher Criticism
to detect anisotropic feature such cosmic strings \citep{starck:jin05}. As weak lensing data may contain filamentary structures, we have also considered such statistical approaches based 
on sparse representations. \\

In section 2, we review the major statistical methods used in the literature to constrain cosmological parameters from weak lensing data. 
Section 3 describes the simulations used in this paper, especially how 2D weak lensing mass maps of five different models have been derived from large statistical samples of 3D N-body simulations of density distribution. 
Section 4 is dedicated to the description of our analysis and we introduce the different statistics we have studied along with the different multiscale transforms investigated. We also present a new statistic that we call Wavelet Peak Counting (WPC). In section 5, we present our results and finally, section 6 et section 7 present a discussion and summaries our conclusions.



\section{Weak lensing statistics and cosmological models constraints: state of the art}
\subsection{Two-point statistics}
\label{twopoint}
The most common method for constraining cosmology in weak lensing studies uses two-point statistics of the shear field calculated either in real or Fourier space. In general there is an advantage in using Fourier space statistics such as the power spectrum because the modes are independent. 
The power spectrum $P_{\kappa}(l)$ of the 2D convergence is defined as a function of the modes $l$ by:
\begin{eqnarray}
< \hat{\kappa}(\vec{l}) \hat{\kappa}(\vec{l}')> = (2\pi)^2 \delta (\vec{l}-\vec{l}') P_{\kappa}(l),
\label{cl}
\end{eqnarray}
where hat symbols denotes Fourier transforms, $\delta$ is the delta function, the brackets denote an average over $l$.
$P_{\kappa}(l)$ only depends on $l=|\vec{l}|$ and $\kappa$ is the convergence (i.e. which is proportional to the projected mass distribution).

This power spectrum $P_{\kappa}(l)$ can be expressed in terms of the 3D matter power spectrum $P(k, \chi)$ of the mass fluctuations $\delta \rho / \rho$ and of cosmological parameters :
\begin{eqnarray}
 P_{\kappa}(l) = \frac{9}{16}\left( \frac{H_0}{c} \right)^2 (\Omega_m)^2 \int{d \chi \left[ \frac{g(\chi)}{ar(\chi)}\right]^2 P\left( \frac{l}{r}, \chi \right),}
\label{cl2}
\end{eqnarray}
where $a$ is the expansion parameter, $H_0$ is the Hubble constant and $\Omega_m$ is the matter density parameter.
The correlation properties are more convenient in Fourier space, but for surveys with complicated geometry due to the removal of bright stars and artifacts, the missing data need proper handling \citep[a review of the different existing methods can be found in][]{inpaint:pires08}. Real space statistics are easier to estimate, but require more computational time. 
The following two-point statistics can be related to the underlying 3D matter power spectrum via the 2D convergence power spectrum $P_\kappa(l)$:
\begin{itemize}
\item The shear variance $<\gamma^2>$ :\\
An example of real space two-point statistic is the shear variance, defined as the variance of the average shear $\bar{\gamma}$ evaluated in circular patches of varying radius $\theta_s$. The shear variance $<\gamma^2>$ is related to the power spectrum $P_{\kappa}(l)$ of the 2D convergence by :
 
\begin{eqnarray}
<\gamma^2> = \int \frac{dl}{2 \pi} l P_{\kappa}(l) \frac{J_1^2(l \theta_s)}{(l \theta_s)^2},
\label{var}
\end{eqnarray}
where $J_n$ is a Bessel function of order $n$.
This shear variance has been used in many weak lensing analysis to constrain cosmological parameters \citep{twopoint:maoli01,twopoint:hoekstra06,twopoint:fu08}.

\item The shear two-point correlation function :\\
The shear two-point correlation function is currently the most used statistic because it is easy to implement and can be estimated even for complex geometry. It is defined as follow :
\begin{eqnarray}
\xi_{i,j}(\theta) = <\gamma_i(\vec{\theta}') \gamma_j(\vec{\theta}' + \vec{\theta})>,
\label{xi}
\end{eqnarray}

where $i, j = 1, 2$ and the averaging is done over pairs of galaxies separated by angle $\theta =|\vec{\theta|}$. By parity $\xi_{1,2} = \xi_{2,1} = 0$ and by isotropy $\xi_{1,1}$ and $\xi_{2,2}$ are functions only of $\theta$.
The shear two-point correlation functions can be related to the 2D convergence power spectrum by : 
\begin{eqnarray}
\xi_+(\theta) = \xi_{1,1}(\theta) + \xi_{2,2}(\theta) = \int_0^\infty \frac{dl}{2 \pi} l P_{\kappa}(l) J_0 (l \theta),
\label{xi2}
\end{eqnarray}
These two-point correlation functions are the most popular statistical tools and have been used in the most recent weak lensing analysis
\citep{twopoint:benjamin07,twopoint:hoekstra06,twopoint:fu08}.

\item The variance of the aperture mass $M_{ap}$ :\\
The aperture mass statistic has been introduced by \cite{map:schneider98}. 
It defined a class of statistics referred to \emph{aperture masses} associated with compensated filters. 
Several forms of filters have been suggested which trade locality in real space with locality in Fourier space.
Considering the filter defined in \cite{map:schneider96} with a cutoff at some scale $\theta_s$.
The variance of the aperture mass can be expressed as a function of the 2D convergence power spectrum by :
\begin{eqnarray}
<M_{ap}^2(\theta_s)> = \int \frac{dl}{2 \pi} l P_{\kappa}(l) \frac{576 J_4^2(l \theta_s)}{(l \theta_s)^4},
\label{xi3}
\end{eqnarray}
This statistic has been used in \cite{twopoint:vanwaerbeke02,twopoint:semboloni06,twopoint:hoekstra06,twopoint:fu08}.  
\end{itemize}

Two-point cosmic shear measurements yield interesting constraints on the amplitude of the matter power spectrum $\sigma_8$ on which the lensing strongly depends. But, when deriving joint constraints, a degeneracy exists between $\sigma_8$ and $\Omega_m$ \citep[see e.g.][]{twopoint:maoli01,wlens:hoekstra02,twopoint:semboloni06}. This degeneracy between $\sigma_8$ and $\Omega_m$ is typical of cosmic shear measurements using only two-point statistics.


Two-point statistics are insufficient to characterize non-Gaussian features. Non-Gaussianity produced by the non-linear evolution of the Universe is of great importance for the understanding of the physics of the Universe, and it may help to better discriminate the cosmological models.

\subsection{Non-Gaussian statistics}
\label{ng}
In the standard structure formation model, initial random fluctuations are amplified by non-linear gravitational instability to produce a final distribution of mass which is highly non-Gaussian.
The weak lensing field is thus highly non-Gaussian. On small scales, we can observe structures like galaxies and clusters of galaxies and on larger scales, we observe some filament structures. Detecting these non-Gaussian features in weak lensing mass maps can be very useful to constrain the cosmological model parameters.

The three-point correlation function $\xi_{i,j,k}$ is the lowest-order statistics which can be used to detect non-Gaussianity.
\begin{eqnarray}
\xi_{i,j,k}(\theta) = <\kappa(\vec{\theta_1}) \kappa(\vec{\theta_2}) \kappa(\vec{\theta_3})>,
\label{xi4}
\end{eqnarray}

In Fourier space it is called bispectrum and only depends on distances $|\vec{l_1}|$,  $|\vec{l_2}|$ and $|\vec{l_3}|$:
\begin{eqnarray}
B(|\vec{l_1}|, |\vec{l_2}|, |\vec{l_3}|) &\propto& <\hat{\kappa}(|\vec{l_1}|) \hat{\kappa}(|\vec{l_2}|) \hat{\kappa}^{*}(|\vec{l_3}|)>.
\label{eq:correl3_3}
\end{eqnarray}

It has been shown that tighter constraints can be obtained with the three-point correlation function \citep{threepoint:bernardeau97,threepoint:cooray01,threepoint:bernardeau03,threepoint:takada03,threepoint:takada04,threepoint:schneider03,threepoint:schneider05,threepoint:benabed06}. 

A simpler quantity than the three-point correlation function is provided by measuring the third-order moment (skewness) of the smoothed convergence $\kappa$ \citep{threepoint:bernardeau97} or of the aperture mass $M_{ap}$ \citep{threepoint:jarvis04,threepoint:kilbinger05}.

Another approach to look for non-Gaussianty is to perform a statistical analysis directly on non-Gaussian structures like clusters. 
Galaxy clusters are the largest virialized cosmological structures in the Universe. 
They provide a unique way to focus on non-Gaussianity present at small scales.
One interesting statistic is the {\em peak counting}  that searches  the number of peaks detected on the 2D convergence corresponding to the cluster abundance \citep[see e.g.][]{peak:hamana04}. 

The methods to search for non-Gaussianity in the weak lensing literature mainly concentrate on higher-order correlation function.
For the CMB, skewness and kurtosis of the wavelet coefficients are standard tools as well  to check the CMB Gaussianity \citep{vielva04,starck:sta05_2,vielva06,Wiaux08},
and it was shown that curvelets  \citep{starck:sta02_3} were useful for the detection of anisotropic feature such cosmic string in the CMB \citep{starck:sta03_1,starck:jin05}.\\

In this following, we perform a comparison between most existing methods in order to find the best higher order statistic to constrain cosmological parameters from weak lensing data. 
To explore the effectiveness of a non-Gaussian measure we use a battery of N-Body simulations. By choosing five models whose two-point correlation statistics are degenerate, we study which statistics are able to distinguish between these models.


\section{Simulations of weak lensing mass maps}

\subsection{3D N-body cosmological simulations}
We have run realistic simulated convergence mass maps derived from N-body cosmological simulations using the RAMSES code \citep{code:Teyssier02}.
The cosmological models are taken to be in concordance with the $\Lambda$CDM model. We have limited the model parameters to a realistic range (see Table \ref{model}) choosing five models along the  ($\sigma_8, \Omega_m$)-degeneracy discussed in \S \ref{twopoint}.\\

\begin{table}[htdp]
\begin{center}
\begin{tabular}{|c|c|c|c|c|c|}
\hline
Model & Box & $\Omega_m$ & $\Omega_L$ & $h$  & $\sigma_8$\\
\hline
model 1& 165.8 & 0.23 & 0.77 & 0.594 & 1\\
\hline
model 2& 159.5 & 0.3 & 0.7 & 0.7 & 0.9\\
\hline
model 3& 152.8 & 0.36 & 0.64 & 0.798 & 0.8\\
\hline
model 4& 145.7 & 0.47 & 0.53 & 0.894 & 0.7\\
\hline
model 5& 137.5 & 0.64 & 0.36 & 0.982 & 0.6\\
\hline
\end{tabular}
\end{center}
\caption{Parameters of the five cosmological models that have been chosen along the ($\sigma_8, \Omega_m$)-degeneracy. The box side is in Mpc/h, the simulations have $256^3$ particles, $\Omega_m$ is the matter density parameter, $\Omega_L$ is the dark energy density parameter, $h$ is equal to $H_0/100$ where $H_0$ is the Hubble constant and $\sigma_8$ the amplitude of the matter power spectrum.}
\label{model}
\end{table}

For each of our five models, we run 21 N-Body simulations, each containing $256^3$ particles. We refined the base grid of $256^3$ cells when the local particle number exceeds 10. We further refined similarly each additional levels up to a maximum level of refinement of 6, corresponding to a spatial resolution of 10 kpc.h$^{-1}$. 


\subsection{2D Weak lensing mass map}

In N-body simulations, which are commonly used in cosmology, the dark matter distribution is represented using discrete massive particles.  The simplest way to treat these particles is to map their positions onto a pixelised grid.  In the case of multiple sheet weak lensing, we do this by taking slices through the 3D simulations.  These slices are then projected into 2D mass sheets.  

The effective convergence can subsequently be calculated by stacking a set of these 2D mass sheets along the line of sight, using the lensing efficiency function. This is a procedure that has been used before by \cite{wlens:vale03}, where the effective 2D mass distribution $\kappa_e$ is calculated by integrating the density fluctuation along the line of sight. 
Using the Born approximation which neglects the facts that the light rays don't follow straight lines, the convergence can be numerically expressed by  :
\begin{eqnarray}
\kappa_e \approx \frac{3H_0^2\Omega_m L}{2c^2}\sum_i\frac{\chi_i(\chi_0-\chi_i)}{\chi_0a(\chi_i)}\left(\frac{n_pR^2}{N_ts^2}-\Delta r_{f_i} \right),
\end{eqnarray}
where $H_0$ is the Hubble constant, $\Omega_m$ is the density of matter, $c$ is the speed of light, L is the length of the box, $\chi$ are co-moving distances, with $\chi_0$ being the co-moving distance to the source galaxies. The summation is performed over the $i^{th}$ box. The number of particles associated with a pixel of the simulation is $n_p$, the total number of particles within a simulation is $N_t$ and $s=L_p/L$ where $L_p$ is the length of the plane doing the lensing. $R$ is the size of the 2D maps and $\Delta r{f_i} = \frac{r_2-r_1}{L}$ where $r_1$ and $r_2$ are co-moving distances.



\begin{figure*}[htdp]
\vbox{
\centerline{
\hbox{
\psfig
{figure=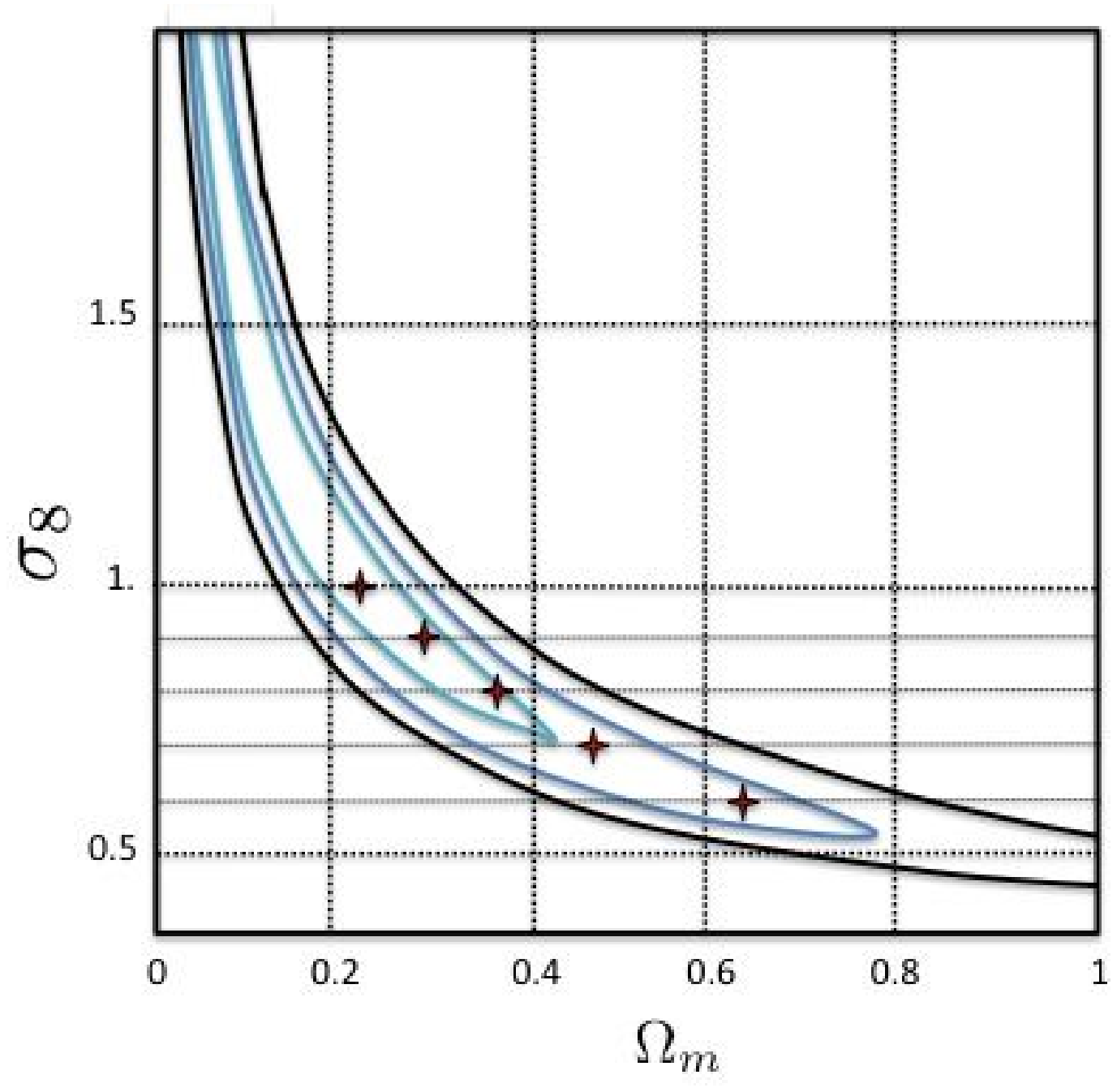,bbllx=5.5cm,bblly=4.cm,bburx=19.5cm,bbury=17.5cm,height=5.5cm,width=5.5cm,clip=}
\hspace{0.2cm}
\psfig{figure=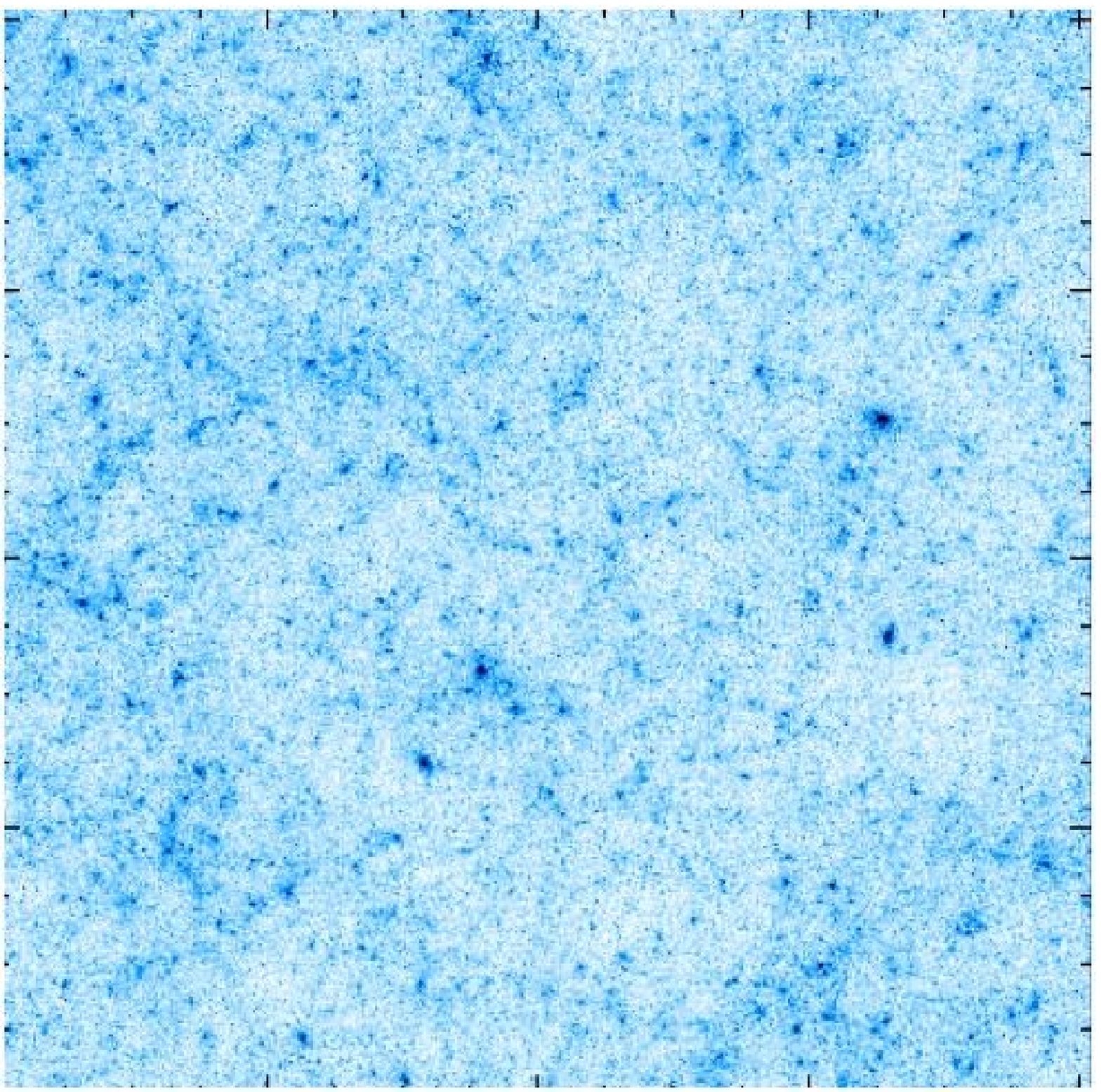,bbllx=0.cm,bblly=0.cm,bburx=17.5cm,bbury=17.5cm,height=5.5cm,width=5.5cm,clip=}
}}
\vspace{0.3cm}
\centerline{
\hbox{
\psfig{figure=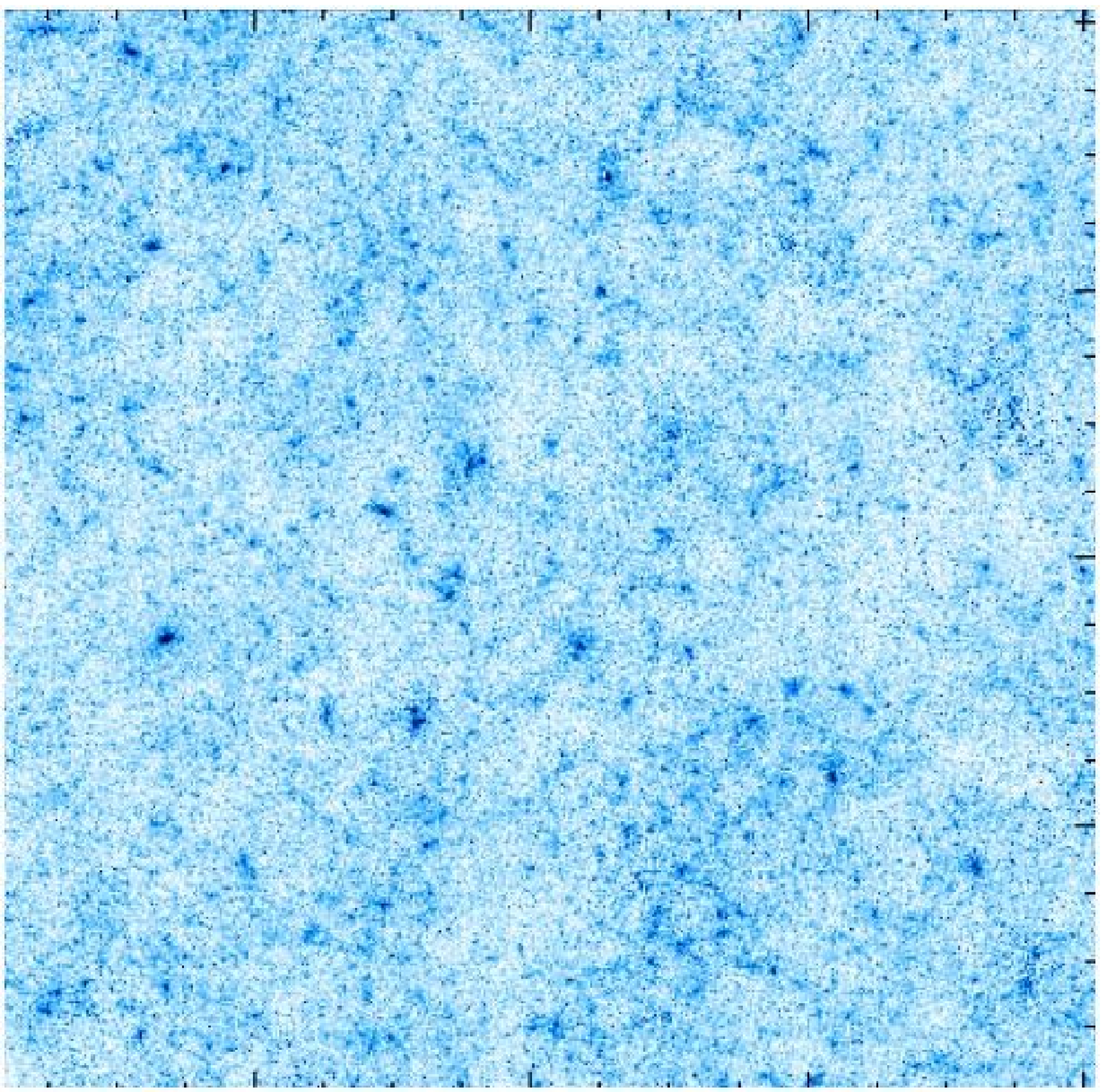,bbllx=0.cm,bblly=0.cm,bburx=17.5cm,bbury=17.5cm,height=5.5cm,width=5.5cm,clip=}
\hspace{0.2cm}
\psfig{figure=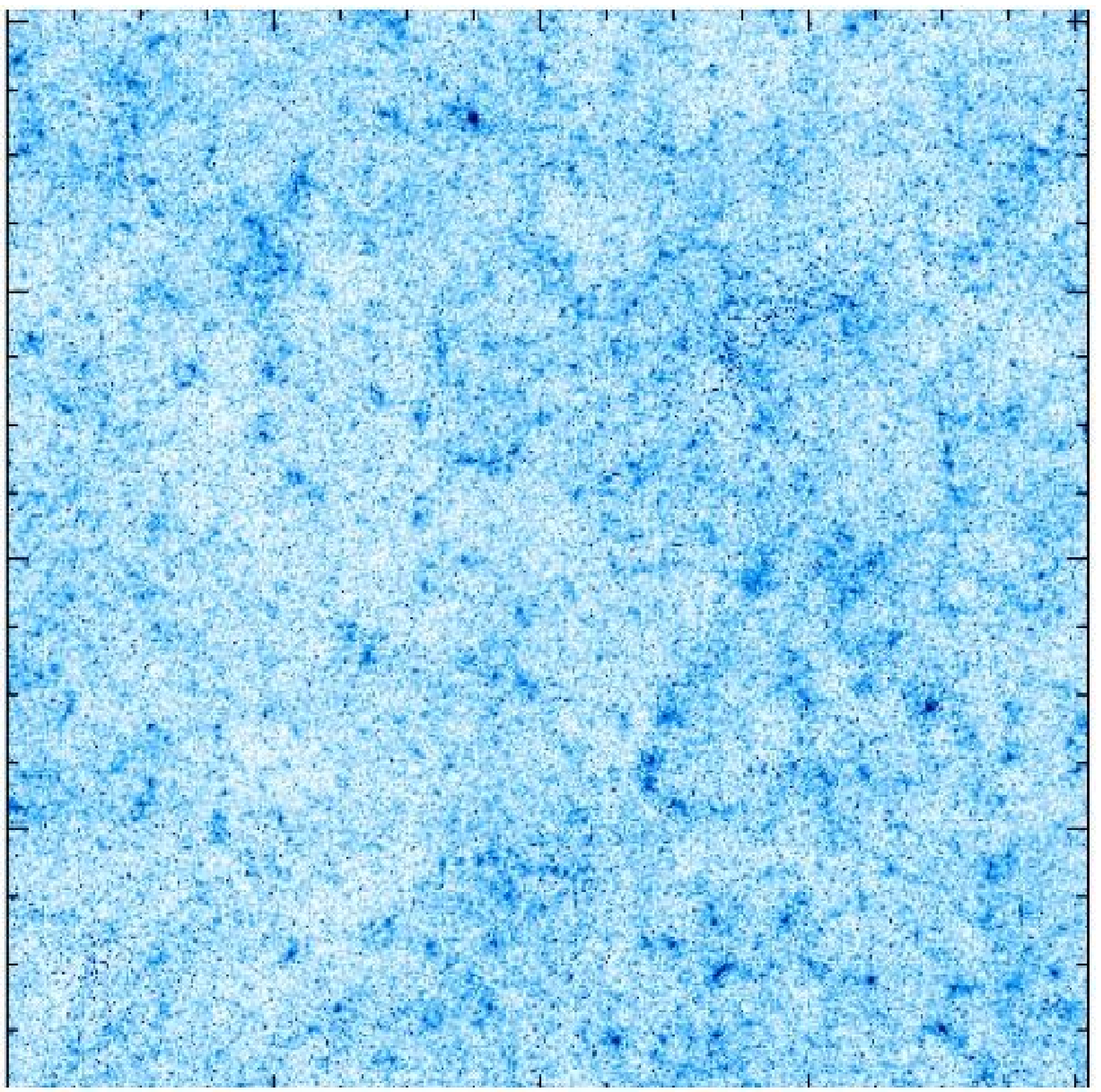,bbllx=0.2cm,bblly=0.cm,bburx=17.2cm,bbury=17.5cm,height=5.5cm,width=5.5cm,clip=}
}}
\vspace{0.3cm}
\centerline{
\hbox{
\psfig{figure=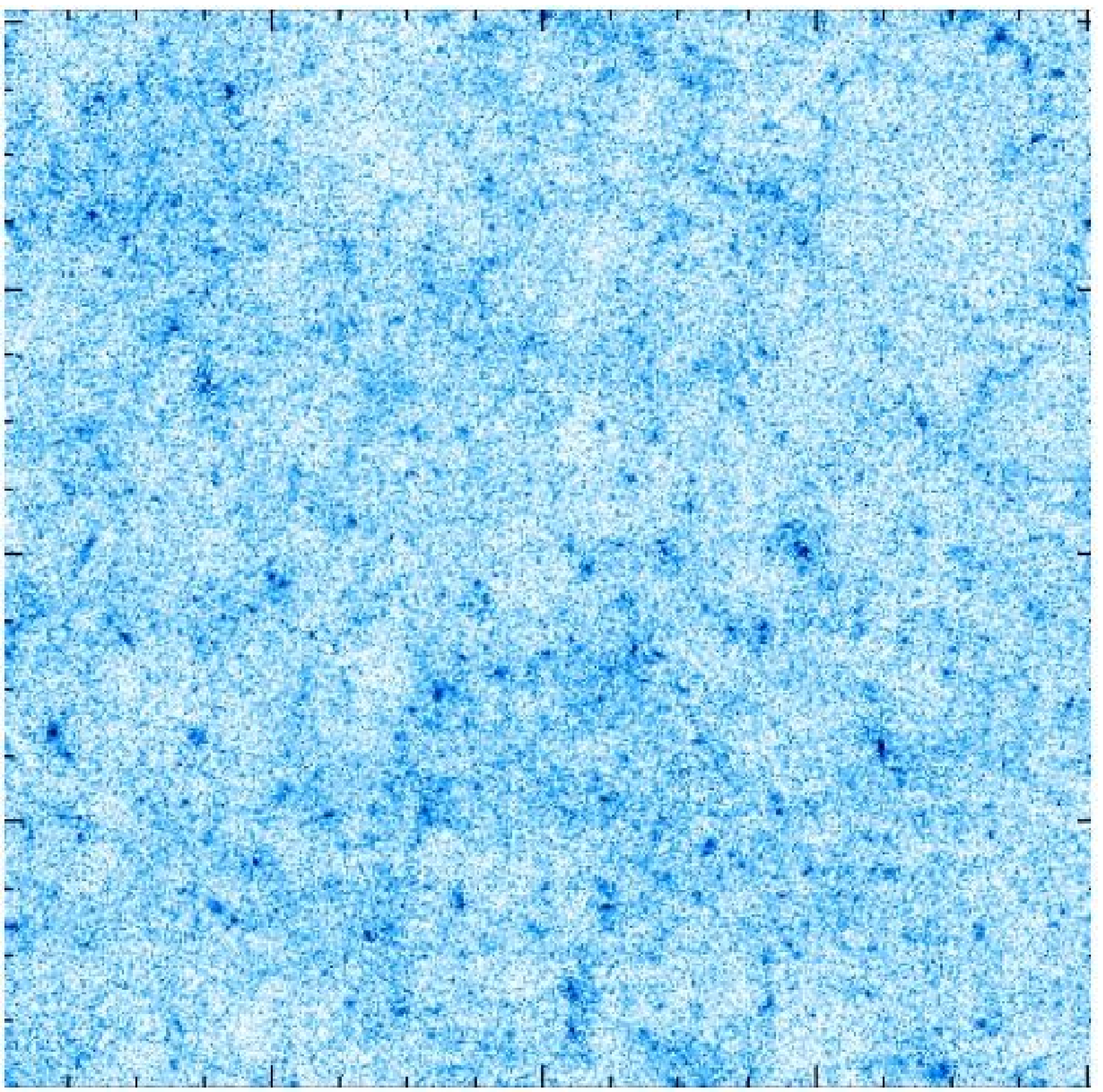,bbllx=0.cm,bblly=0.cm,bburx=17.5cm,bbury=17.1cm,height=5.5cm,width=5.5cm,clip=}
\hspace{0.2cm}
\psfig{figure=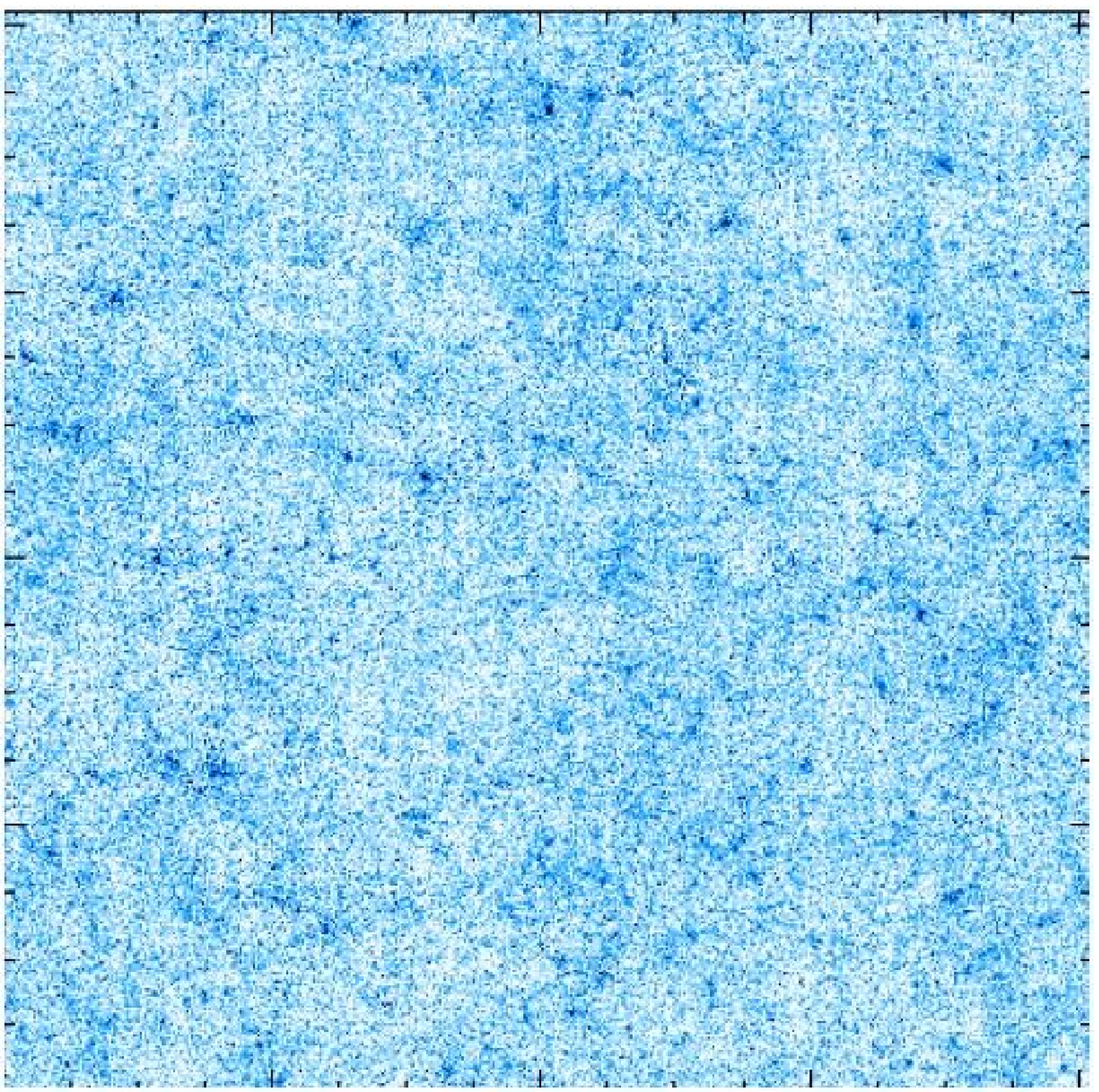,bbllx=0.cm,bblly=0.cm,bburx=17.cm,bbury=17.1cm,height=5.5cm,width=5.5cm,clip=}
}}
}
\caption{Upper left,  the 5 cosmological models along the ($\sigma_8, \Omega_m$)-degeneracy. Upper right, one realization of the convergence $\kappa$ for model 1 ($\sigma_8 = 1$ and $\Omega_m = 0.23$); middle left, for model 2 ($\sigma_8 = 0.9$ and $\Omega_m = 0.3$); middle right, for model 3 ($\sigma_8 = 0.8$ and $\Omega_m = 0.36$); bottom left, for model 4 ($\sigma_8 = 0.7$ and $\Omega_m = 0.47$) and bottom right for model 5 ($\sigma_8 = 0.6$ and $\Omega_m = 0.64$). Each map is 1.975 x 1.975 degrees down-sampled to 512 x 512 pixels}

\label{model0}
\end{figure*}

Using the previous 3D N-body simulations, we have derived 100 realizations for the five models. Fig. \ref{model0} shows one realization of the convergence for each of the 5 models. In all cases, the field is $3.95^{\circ}$ x $3.95^{\circ}$, downsampled to $1024^2$ pixels and we assume that the sources lie at exactly $z=1$. At large scale, the map shows clearly a Gaussian signal. On the contrary, at small scales, the signal is clearly dominated by clumpy structures (dark matter halos) and is therefore highly non-Gaussian. 





\subsection{2D Weak lensing noisy mass map}
\label{sectnoise}
In practice, the observed shear $\gamma_i$ is obtained by averaging over a finite number of galaxies and is therefore noisy. The noise arises both from the measurement errors and from the intrinsic ellipticities dispersion of galaxies.  As a good approximation, we modeled the noise as an uncorrelated Gaussian random field with variance  :
\begin{eqnarray}
\sigma_{noise}^2= \frac{\sigma_{\epsilon}^2}{A.n_g},
\label{noise}
\end{eqnarray}
 where $A$ is the pixel size in arcmin$^2$, $n_g$ the average number of galaxies per arcmin$^2$ and $\sigma_\epsilon$ the rms of the shear distribution. We assume $\sigma_\epsilon \simeq 0.3$ per shear component and $n_g= 100$ gal/arcmin$^2$ as is approximately found for space-based weak lensing surveys \citep{wlens:massey04}. An estimate of the noisy convergence $\kappa_n$ is obtained using the least square estimator defined in \cite{wlens:starck06}.


\section{Cosmological model discrimination framework}
\PS{In this study, in order to find the best statistic to break the ($\sigma_8$, $\Omega_m$)-degeneracy, we are interested in comparing different statistics estimated in different representations using the set of simulated data described in the previous section. 

For this purpose, for each statistic, we need to characterize, the discrimination obtained between each couple of models. The best statistic will be the one that maximizes the discrimination for all the couples of models.}

\subsection{Characterization of the discrimination}
\label{chara}
\begin{figure}[htdp]
\centerline{
\psfig
{figure=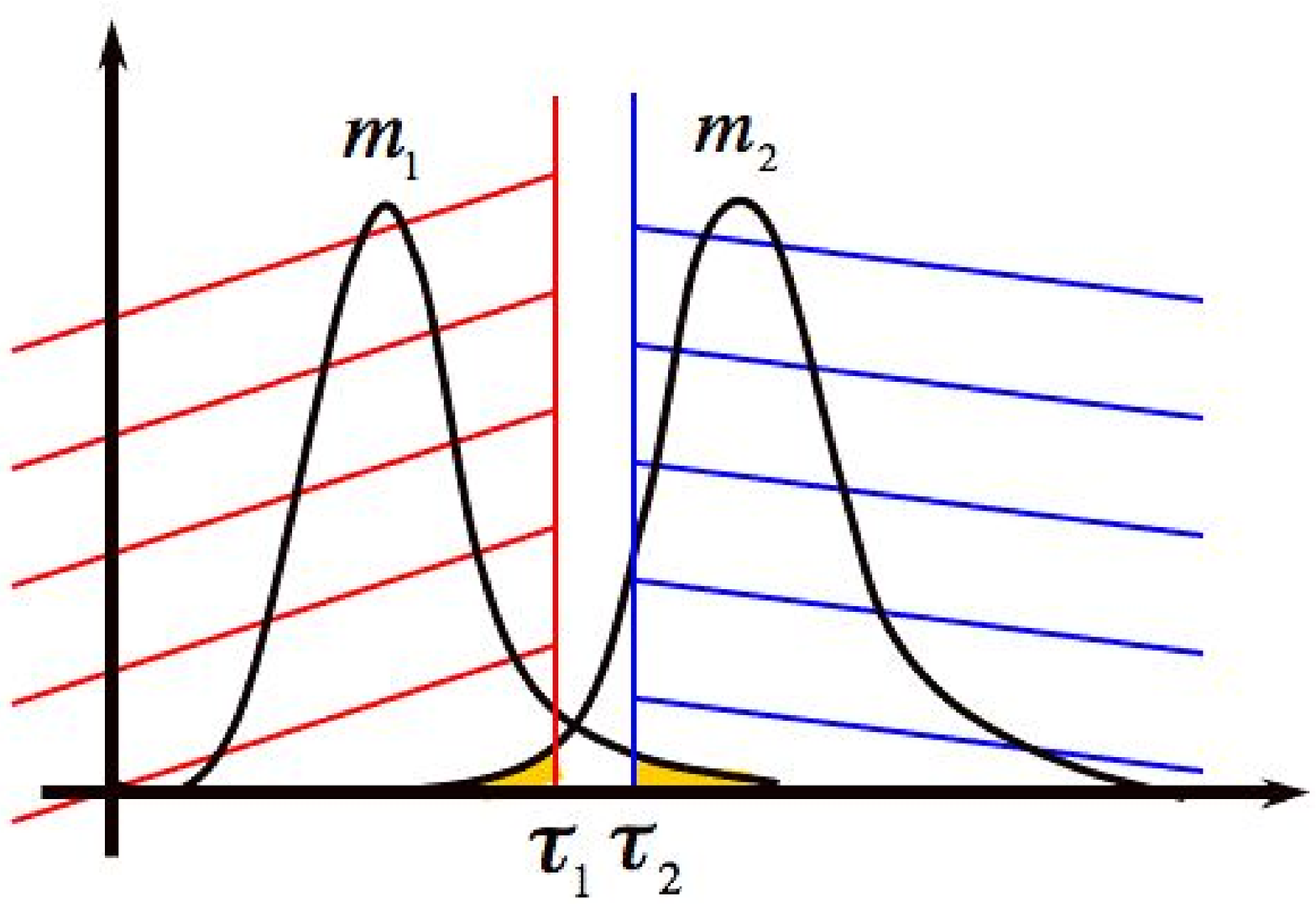,bbllx=5.cm,bblly=3.5cm,bburx=18.5cm,bbury=15.cm,height=5.cm,width=6.cm,clip=}}
\caption{\PS{The following two distributions correspond to the histogram of the values of a given statistic estimated on the 100 realizations of model 1 ($m_1$) and on the 100 realizations of model 2 ($m_2$). The discrimination achieved with this statistic between $m_1$ and $m_2$ is rather good : the two distributions barely overlap. To characterize more quantitatively the discrimination, the FDR method has been used to estimate the thresholds $\mathcal{T}_1$ and  $\mathcal{T}_2$. A false discovery rate ($\alpha$) equal to $0.05$ has been chosen. Then a discrimination efficiency can be derived.}}
\label{distribution}
\end{figure}

\PS{To find the best statistic, we need to characterize quantitatively for each statistic the discrimination between two different models $m_1$ and $m_2$. One way to proceed is to define a discrimination efficiency that expresses the ability of a statistic to discriminate in percentage. Then, we need to define for each individual statistic, two different thresholds (see Fig. \ref{distribution}):\\
- a threshold $\mathcal{T}_1$ : if the estimation of a given statistic in a map $\kappa_i$ is below $\mathcal{T}_1$ the map $\kappa_i$ is classified belonging to the model $m_1$ and not if it is above\\
- a threshold $\mathcal{T}_2$ : if the estimation of a given statistic in a map $\kappa_i$ is above $\mathcal{T}_2 $ the map $\kappa_i$ is classified belonging to the model $m_2$ and not if it is below.
We have used a statistical tool called FDR (False Discovery Rate) introduced by \cite{wlens:benjamini95} to set these two thresholds $(\mathcal{T}_1$ and $\mathcal{T}_2$) correctly (see Appendix A).}

\PS{This FDR method is a competitive tool to set a threshold in an adaptive way without any assumption, given a false discovery rate ($\alpha$). The false discovery rate is given by the proportion of false detections over the total number of detections. The threshold being estimated, we can derive a discrimination efficiency for each statistic. The discrimination efficiency measures the ability of a statistic to discriminate a model from another one by calculating the percentage of detections (true or false) over the total number of samples. It corresponds basically to the part of the distribution that doesn't overlap. The more the distributions overlap, the lower the discrimination efficiency will be.

The Fig. \ref{distribution} represents the dispersion of the values of a given statistic estimated on the 100 realizations of the model 1 (on the left) and on the 100 realizations of the model 2 (on the right). The two distributions barely overlap, it indicates a good discrimination that is to say the two models can easily be separated with this statistic. 
To be more quantitative, a threshold has been set for each distribution to estimate a discrimination efficiency that corresponds to the part of the distribution delimited by the hatched area. 

The formalism of the FDR method ensures that the yellow area delimited by $\mathcal{T}_1$ (resp. $\mathcal{T}_2$) 
that corresponds to  the false detections will be small.}

\subsection{A set of statistical tools }
The first objective of this study is to compare different statistics to find the one that sets tighter constraints on the cosmological parameters. 
The two-point statistics that contain all the information about Gaussian signal leading to the ($\sigma_8,\Omega_m$)-degeneracy, we have opted for statistics currently used to detect non-Gaussianity in order to probe the non-linear process of gravitational clustering. The statistics that we have selected are the following :
\begin{itemize} 
\item Skewness ($S_\kappa$): \\
The skewness is the third-order moment of the convergence $\kappa$ and is a measure of the asymmetry of a distribution.The convergence skewness is primarily due to rare and massive dark matter halos. The distribution will be more or less skewed positively depending on the abundance of rare and massive halos. 
\item Kurtosis ($K_\kappa$): \\
The kurtosis is the fourth-order moment of the convergence $\kappa$ and is a measure of the peakedness of a distribution.  A high kurtosis distribution has a sharper "peak" and flatter "tails", while a low kurtosis distribution has a more rounded peak with wider "shoulders".
\item Bispectrum ($B_\kappa$):\\
Recently, there has been a lot of theoretical works on the three-point correlation function to constrain the cosmological parameters. But the direct computation of three-point correlation function takes too long for our large maps. We have then used its Fourier analog: the Bispectrum that has been introduced \S \ref{ng}. And we have consider the equilateral configuration.
\item Higher Criticism (HC):\\
HC is a recently developed statistic, proposed by \cite{stat:donoho04}. It is a measure of non-Gaussianity obtained by looking at the maximum deviation comparing the sorted $p$-$values$ of a distribution with the sorted $p$-$values$ of a normal distribution. A large HC value implies non-Gaussianity. 
We consider the two different forms of HC (see Appendix B) :\\
- HC$^*$\\
- HC$^+$
\item Peak counting ($P_c$):\\
We now investigate the possibility of using the peak counting to differentiate among cosmological models. By peak counting (or cluster count), we mean the number of halos that we can detect per unit area of the sky (identified as peak above a mass threshold in mass maps).
This cluster count enables to constrain the matter power spectrum normalization $\sigma_8$ for a given $\Omega_m$ \citep[see e.g.][]{peak:bahcall98} and the formalism exist to predict the peak count to a given cosmological model \citep[see e.g.][]{peak:hamana04}. 


\item Wavelet Peak Counting (WPC):\\
We introduce a new statistic that we call Wavelet Peak Counting (WPC). This statistic consists in estimating a cluster count per scale of a wavelet transform. It means we have roughly a cluster count depending on the size of the clusters. We show in the following that the WPC is better than the peak counting to characterize the non-linear structure formation process. 
\end{itemize}

All the statistics are estimated on the 2D convergence $\kappa$ map. In practice the complex geometry of surveys gives the convergence $\kappa$ inferred from the shear field $\gamma_i$ the same complex geometry. But it becomes possible with the inpainting method developed in \cite{inpaint:pires08} to reconstruct a full convergence $\kappa$ map. The solution that is proposed enables to fill-in judiciously masked regions so as to reduce the impact of missing data on the estimation of the power spectrum and of higher-order statistical measures while requiring $O(N \log N)$ operations.

\subsection{Representations}
The second objective of the present study is to compare different transforms to find the sparsest representation of weak lensing data that makes the discrimination easier. Some studies on CMB data have used multiscale methods to detect non-Gaussianity \citep{ng:aghanim99,ng:starck04}. Weak lensing maps exhibit both isotropic and anisotropic features. These kind of features can be better represented in some basis functions. Indeed, a transform is optimal to detect structures which have the same shape of its basis elements. We have thus tested different representations :
\begin{itemize}
\item The Fourier transform
\item The anisotropic bi-orthogonal wavelet transform\\
We expect the bi-orthogonal wavelet transform to be optimal for detecting mildly anisotropic features. \item The isotropic "\`a trous" wavelet transform \\
This wavelet transform is well adapted to the detection of isotropic features such as the clumpy structures (clusters) of the weak lensing data.
\item The ridgelet transform\\
The ridgelet transform has been developed to process images including ridge elements, and so provides a good representation of perfectly straight edges.
\item The curvelet transform\\
The curvelet transform allows to approximate curved singularities with few coefficients and then provides a good representation of curved structures.
\end{itemize}
A description of these transforms is available in Appendix C.
We have to notice that the estimation of a given statistic at each resolution level of a representation can be considered as an individual statistic by itself.

\section{Analysis and results}
\subsection{Treatment of the noise}
As explained previously, the weak lensing mass maps are measured from the distortions of a finite number of background galaxies and therefore suffer from shot noise. Furthermore each galaxy provides only a noisy estimator of the distortion field. We have added the expected level of Gaussian noise (see \S \ref{sectnoise}) to simulations of weak lensing mass maps 
 to obtain simulated noisy mass maps corresponding to space observations. 
Fig. \ref{noise} shows a noiseless simulated mass map (left) and a noisy simulated mass map (right) corresponding to space observations.
 
 \begin{figure*}[htdp]
\centerline{
\hbox{
\psfig{figure=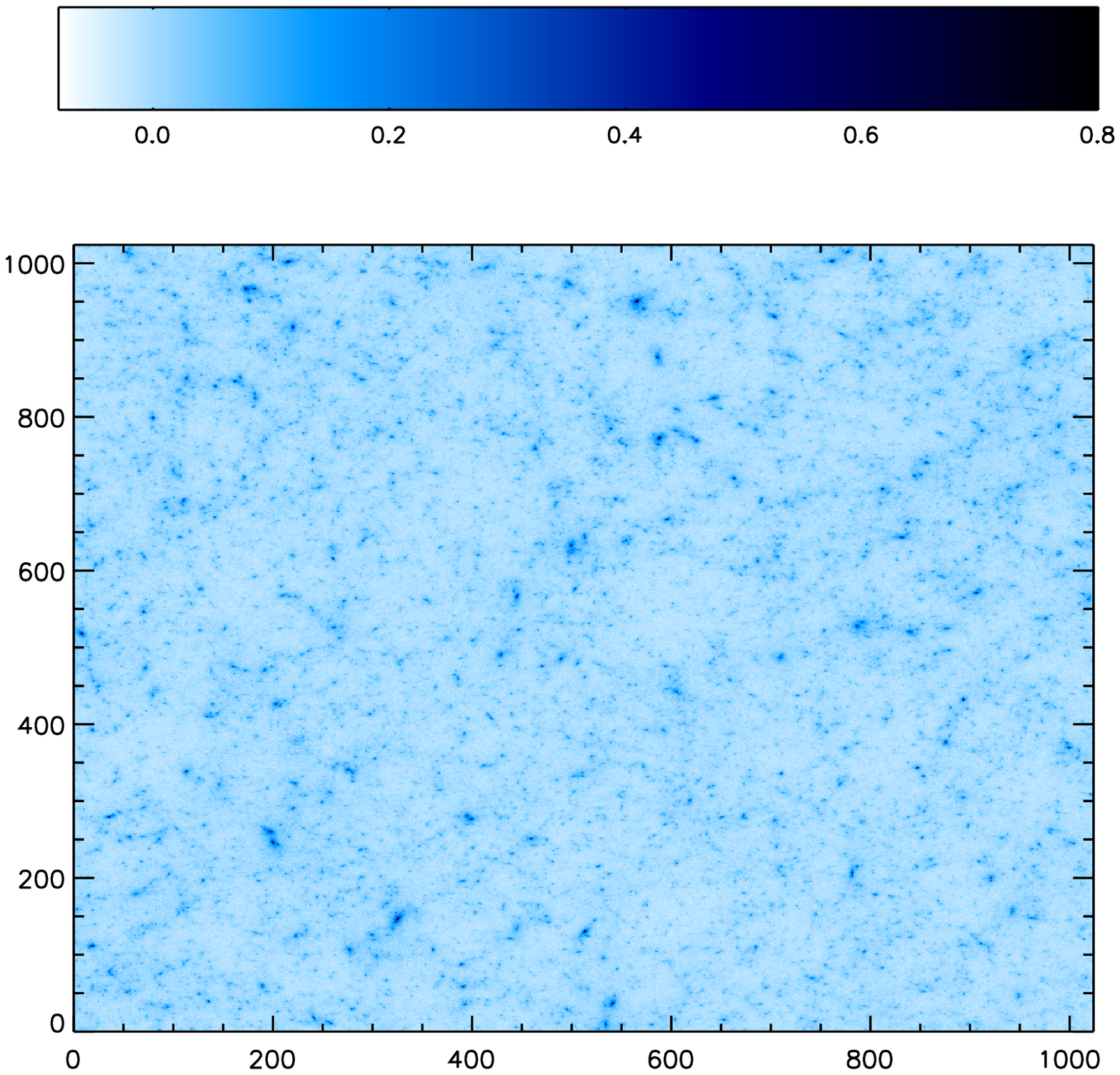,bbllx=4.cm,bblly=6.5cm,bburx=20.cm,bbury=21.cm,height=7.5cm,width=6.5cm,clip=}
\hspace{0.2cm}
\psfig{figure=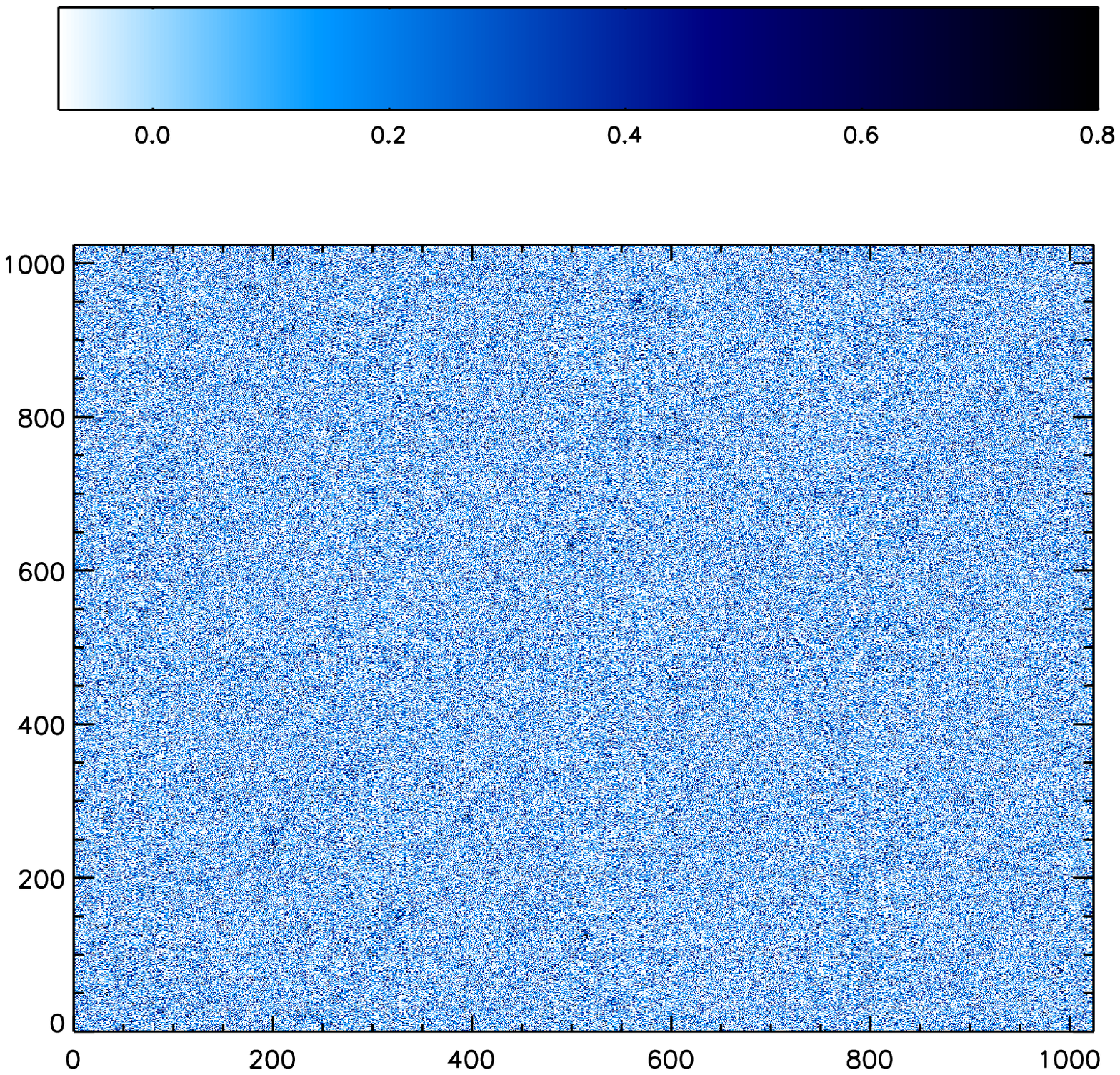,bbllx=4.cm,bblly=6.5cm,bburx=20.cm,bbury=21.cm,height=7.5cm,width=6.5cm,clip=}
}}
\caption{ Left: noiseless simulated mass map and right: simulated noisy mass map that we should obtain in space observations. The field is 1.975$^{\circ}$ x 1.975$^{\circ}$.}
\label{noise}
\end{figure*}
The noise has an impact on the estimated statistics and therefore needs to be considered.

 \subsubsection{No filtering}
We begin by applying the different transformations directly to the noisy data 
and for each representation, we have estimated the statistics described in the previous section, 
with the exception of cluster count and $WPC$ which required filtering. 

We expect that the noise will make the third and fourth order statistics tend to zero. 
Indeed the more noisy the data are, the more the distribution will look like a Gausssian.

 \subsubsection{Gaussian filtering}
 As a second step, we have performed a Gaussian filtering that is obtained by convolving the noisy mass maps $\kappa_n$ with a Gaussian window $G$ with a standard deviation $\sigma_G$ :
 \begin{eqnarray}
\kappa_G = G * \kappa_n.
\label{Gaussian}
\end{eqnarray}

\begin{figure*}[htdp]
\centerline{
\hbox{
\psfig{figure=immmm1.ps,bbllx=4.cm,bblly=6.5cm,bburx=20.cm,bbury=19.cm,height=6.5cm,width=6.5cm,clip=}
\hspace{0.2cm}
\psfig{figure=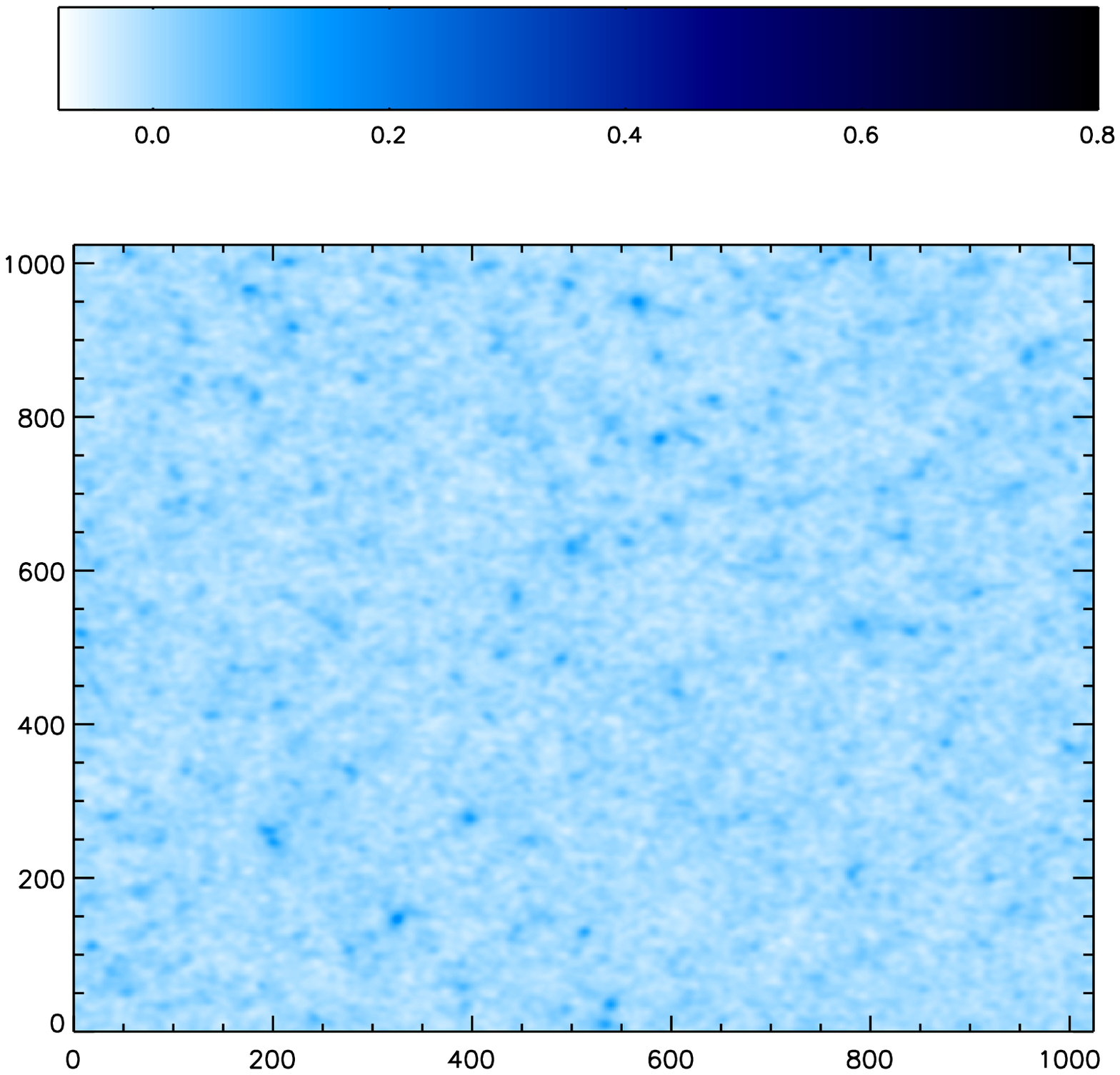,bbllx=4.cm,bblly=6.5cm,bburx=20.cm,bbury=19.cm,height=6.5cm,width=6.5cm,clip=}
}}
\caption{Left, noiseless simulated mass map, and right, filtered mass map by convolution with a Gaussian kernel. The field is 1.975$^{\circ}$ x 1.975$^{\circ}$.}
\label{gaus}
\end{figure*}

Fig. \ref{gaus} shows on the left, the original map without noise and on the right, the result obtained by Gaussian filtering of the noisy mass map displayed in Fig. \ref{noise} (right). The quality of the filtering depends strongly on the value of $\sigma_G$. For the simulations, an optimal value is about $0.9$ arcmin). 
We then computed all our statistics on this new set of filtered mass maps.

\subsubsection{MRLens filtering}
  
Finally, we used a method of non-linear filtering based on the wavelet representation : the MRLens filtering proposed by \cite{wlens:starck06}. 
\PS{The MRLens filtering is based on the Bayesian ideas that provides the means to incorporate prior knowledge in data analysis. 
Choosing the prior is one of the most critical aspects of Bayesian analysis. Here a multiscale entropy prior is used.
A full description of the method is given in the Appendix D.
The MRLens software that we used is available at the following address : http://www-irfu.cea.fr/Ast/878.html. 


It was shown in \cite{wlens:starck06} that this method outperforms several standard techniques (Gaussian filtering, Wiener filtering, MEM filtering,etc.).
Fig. \ref{filtering} shows on the left, the original map without noise and on the right, the result of a FDR filtering on the noisy mass map displayed Fig. \ref{noise} (right). The visual aspect indicates that many clusters are reconstructed and the intensity of the peaks are well recovered. 

\begin{figure*}[htdp]
\centerline{
\hbox{
\psfig{figure=immmm1.ps,bbllx=4.cm,bblly=6.5cm,bburx=20.cm,bbury=19.cm,height=6.5cm,width=6.5cm,clip=}
\hspace{0.2cm}
\psfig{figure=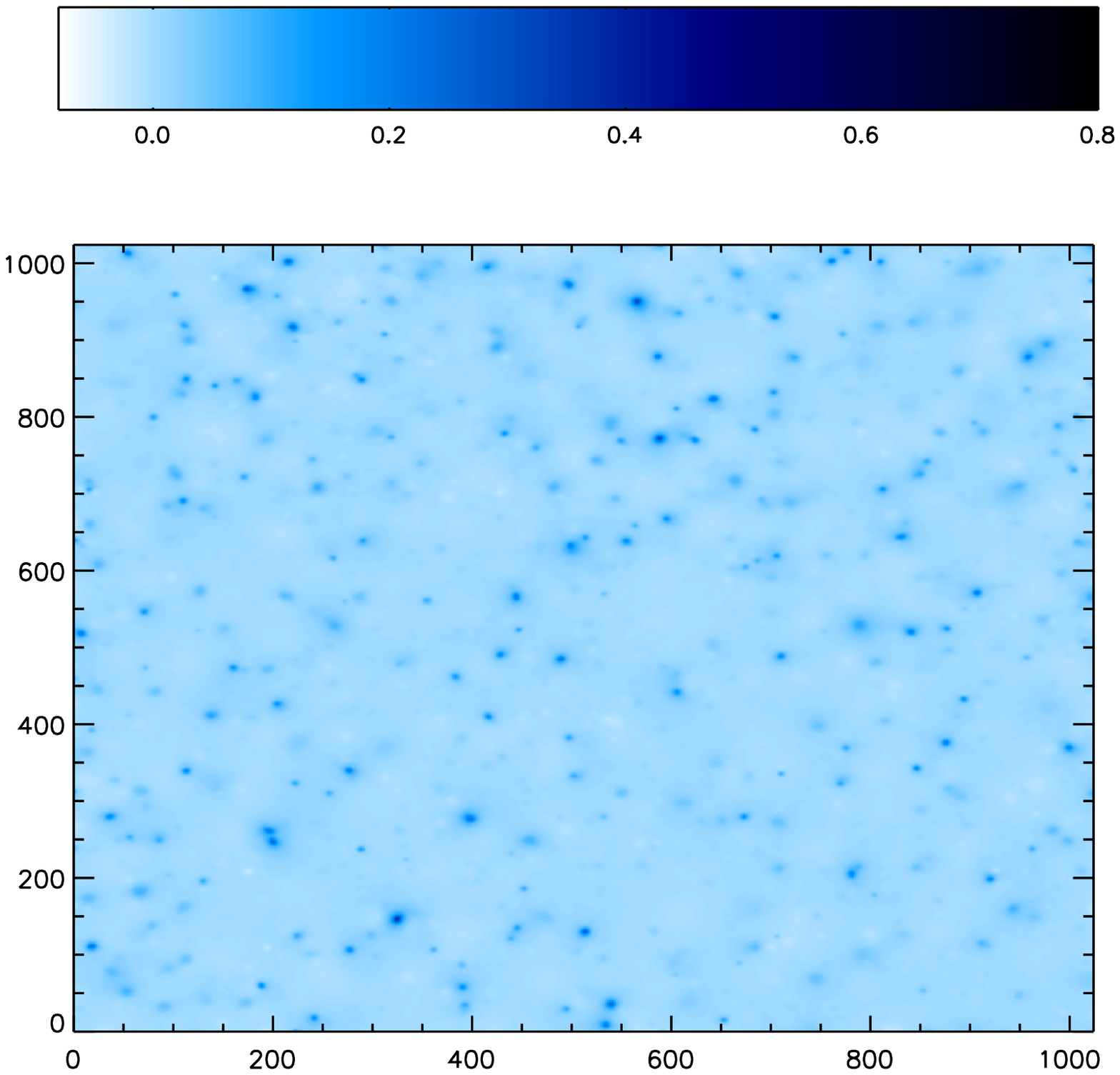,bbllx=4.cm,bblly=6.5cm,bburx=20.cm,bbury=19.cm,height=6.5cm,width=6.5cm,clip=}
}}
\caption{Left, noiseless simulated mass map, and right, filtered mass map by the FDR multiscale entropy filtering. The field is 1.975$^{\circ}$ x 1.975$^{\circ}$.}
\label{filtering}
\end{figure*}

As done before, all the statistics have been estimated on these MRLens filtered mass maps.}
By reconstructing essentially the clusters, we can anticipate that this MRLens method will help statistics like peak counting or WPC more than statistics that focus on the background.
 
\subsection{Results}
\subsubsection{The discrimination methodology}
\PS{As explained section \ref{chara}, for each statistic described in the previous section, we can derive a discrimination efficiency between each two models out of the full set of 5 models. This values are given Table \ref{disc_skew_2}, Table \ref{disc_gaus_2} and Table \ref{disc_wl_2} for three different statistics. 

A mean discrimination efficiency for each individual statistic can be estimated by averaging the discrimination efficiency across all the pairs of models.
For statistics estimated in multiscale representations, the mean discrimination efficiency is calculated for each scale and it is the better one that is considered.

Tables \ref{disc_noise}, \ref{disc_gaus} and \ref{disc_filter} show the mean discrimination efficiency reached by a given statistic and a given transform, respectively for (i) unfiltered mass maps, (ii) Gaussian filtered mass maps and (iii) MRLens mass maps. 

The mean discrimination efficiency of the Table \ref{disc_skew_2} is about $40\%$ and corresponds to the value at position (1,3) of Table \ref{disc_noise} (i.e. the skewness of wavelet coefficients). And, the mean discrimination efficiency of Table \ref{disc_wl_2} is about $97\%$ and corresponds to the value at position (3,5) of Table \ref{disc_filter}. 

A full discrimination between the five models is obtained if the mean discrimination efficiency is about $100\%$. If it is between $80\%$ and $100\%$ the discrimination is possible except for adjacent models. Less than $40\%$ there is no discrimination possible even for distant models.}

\subsubsection{Discrimination in noisy mass maps}

The mean discrimination efficiency obtained for unfiltered mass maps is displayed in Table \ref{disc_noise}.
The peak counting and WPC differ from the others. They can not be computed on unfiltered mass maps because the clusters can not be extracted from noisy mass maps. Another point is that the bispectrum by definition can only be estimated in the Fourier domain.

Without filtering the results are poor and no discrimination can be achieved in Direct space. Indeed, the Signal-to-Noise Ratio is weak as can be seen in Fig. \ref{noise} (right). The non-Gaussian signal is hidden by the Gaussian noise.

The different transforms appear to be inefficient at bring out the non-Gaussianity features, except, the Isotropic Wavelet Transform (see Table \ref{disc_noise}) which performs rather well, whatever the statistic. This is likely because it is optimal to detect clusters. Indeed, the clusters are a direct probe of non-Gaussianity and by concentrating the cluster information, the isotropic wavelet transform seems to be a better representation for non-Gaussianity.
\begin{table*}[htdp]
\begin{center}
\begin{tabular}{|c|c|c|c|c|c|c|c|}
\hline
Basis/Statistics &  $S$ & $K$ & $HC_n^*$  & $HC_n^+$  & $B_l$ & $P_c$& $WPC$\\
\hline
Direct space & 21.0 & 1.6 & 1.8 & 1.9 & x & x& x \\
\hline
Fourier space &  0.7 & 10.7 & 3.0 & 21.3 & 5.75 & x& x \\
\hline
Isotropic Wavelet Transform & 40.5 & 29.3 & 24.9 & 34.0 & x & x& x\\
\hline
Bi-orthogonal Wavelet Transform & 7.3 & 10.8 & 4.1 & 2.9& x & x& x\\
\hline
Ridgelet Transform (block = 16) & 6.2 & 3.7 & 7.2 & 5.6 & x & x& x\\
\hline
Ridgelet Transform (block = 32) & 8.3 & 15.2 & 9.5 & 20.4& x  & x& x\\
\hline
Fast Curvelet Transform & 1.1 & 4.2 & 2.6 & 4.2 & x & x& x\\
\hline
\end{tabular}
\end{center}
\caption{Mean discrimination efficiencies (in percent) achieved on noisy mass maps  with a False Discovery Rate $\alpha = 0.05$.}
\label{disc_noise}
\end{table*}

\begin{figure}[htdp]
\centerline{
\psfig
{figure=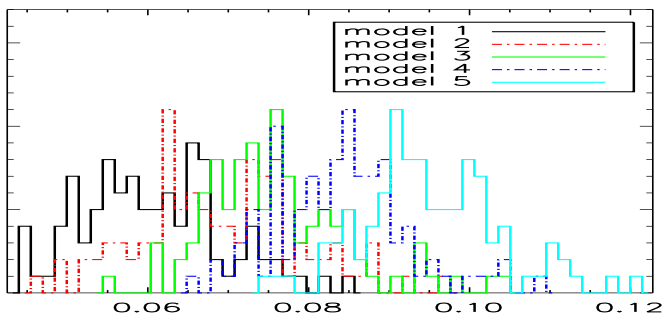,bbllx=5.5cm,bblly=5.5cm,bburx=19.cm,bbury=21.5cm,height=4.5cm,width=9.cm,clip=}}
\caption{\PS{Distribution of the skewness calculated from the second scale of an Isotropic Wavelet Transform on the simulated realizations of the 5 models. It illustrates the results of the Table 3. No discrimination is possible except between the farthest models (i.e. between model 1 and model 5)}}
\label{discri1}
\end{figure}

The skewness in the Wavelet transform representation appears to be the best statistic in unfiltered mass maps. Table \ref{disc_skew_2} shows the discrimination efficiency obtained with the skewness at the second scale of an Isotropic Wavelet Transform being the scale that reaches the best discrimination. We can see that the discrimination is only achieved between the farthest models, which is quite poor. \PS{It is illustrated on Fig. \ref{discri1} where you can see that the 5 distributions are overlapping.}
Some groups have already used the skewness aperture mass statistic to try to break the $\sigma_8$-$\Omega_m$ degeneracy, \cite[see e.g.][]{threepoint:kilbinger05,threepoint:jarvis04}. This processing consists of convolving the noisy signal by Gaussian windows with different scales and is quite similar to an Isotropic Wavelet Transform. They showed that by combining the second and third-order statistics, the degeneracy can be diminished but not broken.

\begin{table*}[htdp]
\begin{center}
\begin{tabular}{|c|c|c|c|c|c|}
\hline
&  model 1 & model 2 &model 3  & model 4 & model 5\\
\hline
model 1 & x & 4 & 55 & 86 & 99 \\
\hline
model 2 & 6 & x & 8 & 52 & 88  \\
\hline
model 3 & 23 & 3 & x & 6 & 77 \\
\hline
model 4 & 81 & 7 & 2 & x & 13 \\
\hline
model 5 & 99 & 81 & 12 & 9 & x\\
\hline
\end{tabular}
\end{center}
\caption{Discrimination efficiencies (in percent) achieved on unfiltered mass maps with the skewness estimated at the second scale of an Isotropic Wavelet Transform.}
\label{disc_skew_2}
\end{table*}

\subsubsection{Discrimination in Gaussian filtered mass maps}

To increase the Signal-to-Noise Ratio, we have applied a Gaussian filtering to the noisy simulated mass maps. Table \ref{disc_gaus} shows the results.

\begin{table*}[htdp]
\begin{center}
\begin{tabular}{|c|c|c|c|c|c|c|c|}
\hline
Basis/Statistics &  $S$ & $K$ & $HC_n^*$  & $HC_n^+$ & $B_l$ & $P_c$& $WPC$ \\
\hline
Direct space & 39.2 & 34.3 & 28.8 & 40.2 & x & 79.3&x\\
\hline
Fourier space & 2.2 & 2.2 & 3.8 &  2.9 & 6.75  & x&x\\
\hline
Isotropic Wavelet Transform & 31.9 & 29.5 & 24. & 38.4 & x &  x&79.2\\
\hline
Bi-orthogonal Wavelet Transform & 12.2 & 10.5 & 5.1 & 6.8 & x & x&x\\
\hline
Ridgelet Transform (block = 16) & 7.4 & 20.2 & 22.7 & 35.3 & x & x&x\\
\hline
Ridgelet Transform (block = 32) & 2.5 & 7.3 & 4. & 9.1 & x & x&x\\
\hline
Fast Curvelet Transform & 1.3 & 4.7 & 4.8 &  8.3 & x & x&x\\
\hline
\end{tabular}
\end{center}
\caption{Mean discrimination efficiencies (in percent) achieved on Gaussian filtered mass maps  with a False Discovery Rate $\alpha = 0.05$.}
\label{disc_gaus}
\end{table*}

After Gaussian filtering, the noise is removed but the structures are over-smoothed. Except for the direct space where the results are clearly improved by the noise removal, the results after a Gaussian filtering are quite similar. Some statistics are a bit improved by the noise removal while others become worse.

\begin{figure}[htdp]
\centerline{
\psfig
{figure=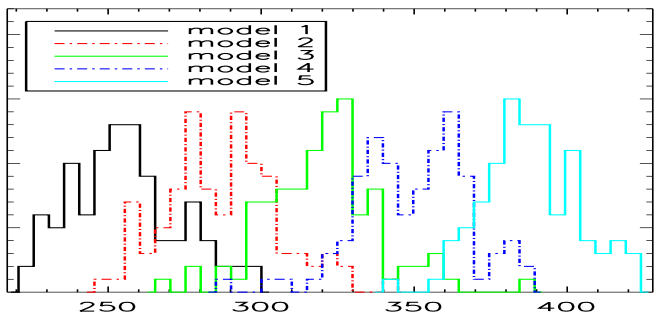,bbllx=5.5cm,bblly=5.5cm,bburx=19.cm,bbury=21.5cm,height=4.5cm,width=9.cm,clip=}}
\caption{\PS{Distribution of the peak counting estimated directly on the simulated realizations of the 5 models. It illustrates the results of the Table 5. The discrimination is possible except between adjacent models (that is to say between model 1 and model 2, model 2 and model 3, model 3 and model 4, model 4 and model 5).}}
\label{discri2}
\end{figure}

By contrast, the peak counting and WPC that can now be estimated on these filtered mass maps perform well. Table \ref{disc_gaus_2} shows the discrimination efficiency obtained with peak counting estimated on Gaussian filtered mass maps. We can see that except for adjacent models, discrimination is now possible. \PS{We can verify this results by looking at the 5 distributions displayed Fig. \ref{discri2}. Indeed, the distributions barely overlap  for no adjacent models.}

The ability of weak lensing cluster count (peak counting) to discriminate between the five different models chosen along the degeneracy can be explained. Considering that the dark matter lies at $z \simeq 0.5$, which is the maximum lensing efficiency for background galaxies at $z=1$, and assuming a constant dark energy model, the number density of massive clusters in the weak lensing mass maps is sensitive to both the amplitude of the mass fluctuations $\sigma_8$ and the matter density parameter $\Omega_m$, both at $z=0$. If instead of $\sigma_8$, we consider $\sigma_8^{z \simeq 0.5}$ that corresponds to the amplitude of the fluctuations of the projected weak lensing mass map. The $\sigma_8^{z \simeq 0.5}$ is now a constant along the five models because the five corresponding weak lensing power spectrum are undistinguishable. 
This leaves the $\Omega_m$ parameter that drives the structure formation \citep[see e.g.][]{peak:bahcall98}. A small $\Omega_m$ makes the structures form earlier and a large $\Omega_m$  makes the structures form later. Then, with a small $\Omega_m$, the abundance of massive clusters at $z \simeq 0.5$ are more significant (see Fig. \ref{model0} upper right) than for a large $\Omega_m$ 
The cluster count can then be used to discriminate cosmological model. The massive cluster abundance has already been used  to probe $\Omega_m$ \citep[see e.g.][]{peak:bahcall98} .

\begin{table*}[htdp]
\begin{center}
\begin{tabular}{|c|c|c|c|c|c|}
\hline
 &  model 1 & model 2 &model 3  & model 4 & model 5\\
\hline
model 1 & x & 50 & 95 & 100 & 100 \\
\hline
model 2 & 46 & x & 43 & 96& 100  \\
\hline
model 3 & 94 & 8. & x & 63. & 100 \\
\hline
model 4 & 100 & 97 & 50 & x & 78\\
\hline
model 5 & 100 & 100 & 99 & 63 & x\\
\hline
\end{tabular}
\end{center}
\caption{Discrimination efficiencies (in percent) achieved on Gaussian filtered mass maps with the peak counting statistic on direct space given a False Discovery Rate $\alpha = 0.05$.}
\label{disc_gaus_2}
\end{table*}

\subsubsection{Discrimination in MRLens filtered mass maps}

Table \ref{disc_filter} shows the results obtained in the case where the MRLens filtering scheme is applied to the noisy simulated mass maps.

\begin{table*}[htdp]
\begin{center}
\begin{tabular}{|c|c|c|c|c|c|c|c|}
\hline
Basis/Statistics &  $S$ & $K$ & $HC_n^*$  & $HC_n^+$ & $B_l$& $P_c$&$WPC$ \\
\hline
Direct space & 65.2 & 54.5 & 72.9 & 8.4 & x & 93.2& x\\
\hline
Fourier space & 3.2 & 3.4 & 30.0 & 38.4 & 9.95  & x & x\\
\hline
Isotropic Wavelet Transform & 81.7 & 68.1 & 74.8 & 74.1 & x  & x& 97.3\\
\hline
Bi-orthogonal Wavelet Transform & 52.2 & 67.3 & 29.7 & 65.0 & x  & x& x\\
\hline
Ridgelet Transform (block = 16) & 45.1 & 54.5 & 77.1 & 37.2 & x  & x& x\\
\hline
Ridgelet Transform (block = 32) & 61.5 & 52.6 & 70.3 & 69.0 & x  & x& x\\
\hline
Fast Curvelet Transform & 25. & 65.5 & 68.8 & 63.3 & x  & x& x\\ 
\hline
\end{tabular}
\end{center}
\caption{Mean discrimination efficiencies (in percent) achieved on MRLens filtered mass maps with a False Discovery Rate $\alpha = 0.05$.}
\label{disc_filter}
\end{table*}

After MRLens filtering, the sensitivity of all transforms is greatly improved. But, the Isotropic Wavelet Transform remains the best transform certainly helped by the MRLens filtering that uses this transform. 

The best statistics remains the peak counting, which is also helped by the MRLens filtering that reconstructs essentially the clusters. But the others statistics also achieve quite good results in these MRLens filtered mass maps compared with the one with Gaussian filtered mass maps. This is likely because the MRLens filtering by favoring the clusters reconstruction helps all the statistics that look for non-Gaussianity.\\

\begin{figure}[htdp]
\centerline{
\psfig
{figure=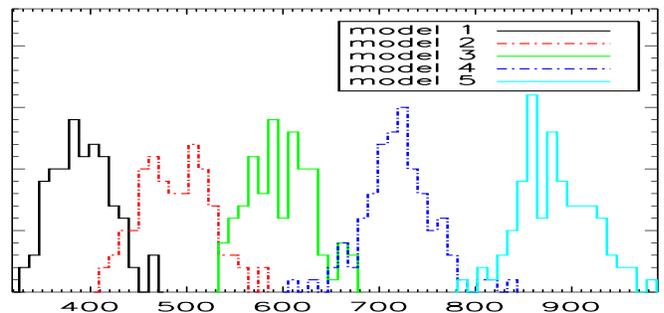,bbllx=5.5cm,bblly=5.5cm,bburx=19.cm,bbury=21.5cm,height=4.5cm,width=9.cm,clip=}}
\caption{\PS{Distribution of the Wavelet Peak Counting estimated at the third scale of an Isotropic Wavelet Transform on the simulated realizations of the 5 models. It illustrates the results of the Table 7. We obtain a good discrimination even for adjacent models. }}
\label{discri3}
\end{figure}

The best result is obtained with WPC on the third scale of the Isotropic wavelet transform see Table \ref{disc_wl_2}. \PS{Fig. \ref{discri3} shows the 5 distributions that  barely overlap.} This statistic allows us to discriminate between different models even for adjacent models for which the discrimination is challenging.

\begin{table*}[htdp]
\begin{center}
\begin{tabular}{|c|c|c|c|c|c|}
\hline
 &  model 1 & model 2 &model 3  & model 4 & model 5\\
\hline
model 1 & x & 86 & 100 & 100 & 100 \\
\hline
model 2 & 87 & x & 94 & 100 & 100  \\
\hline
model 3 & 100 & 92 & x & 94 & 100 \\
\hline
model 4 & 100 & 100 & 93 & x & 99 \\
\hline
model 5 & 100 & 100 & 100 & 100 & x\\
\hline
\end{tabular}
\end{center}
\caption{Discrimination efficiencies (in percent) achieved on MRLens filtered mass maps with WPC at the third scale of an Isotropic Wavelet Transform.}
\label{disc_wl_2}
\end{table*}

The comparison of these results with the results obtained on noisy mass maps by the skewness in a wavelet representation (Table \ref{disc_skew_2}) show that the accuracy on $\sigma_8$ and $\Omega_m$ is greatly improved using WPC estimated on MRLens filtered mass maps. 

\section{Discussion}
\label{discussion}
As stated earlier, the formalism of the halo model provides a prediction for the number of clusters contained in a given field for a given cosmological model \citep{halo:press74,halo:sheth99,peak:hamana04}.
However, we have to consider that just a fraction of the clusters, present in the sky, will be detected.
It follows that we have to take into account the selection effects coming from the observation quality and the data processing method.
The solution that is currently used consists in modeling the selection effects by estimating the selection function. An analytic model can be done by considering all the selection effects.
An alternative consists in using a Monte Carlo approach that enables us to take into account the entirety of the selection effects that could not be considered in the analytic approach. This study will be done in a future work.

The selection function being specified, the connection with observations and theory is straightforward. The cosmological parameters can thus be estimated from $WPC$.

Having a perfect discrimination between the 5 cosmological models with $WPC$ upper limits error on the cosmological parameters can be given by considering the spacing between two adjacent models. For space observations covering 4 square degrees, the upper limit error in $\sigma_8$ is $8\%$, in the range of $\sigma_8 \in [0.6,1]$. And the upper limit error in $\Omega_m$ is $12\%$, in the range of $\Omega_m \in [0.23,0.64]$. In future work, an accurate estimation of the error should be done.

\section{Conclusion}


Using only two-point statistics to constrain the cosmological model yield to various degeneracies between cosmological parameters such as the ($\sigma_8$-$\Omega_m$)-degeneracy. In this paper, we survey a range of non-Gaussian statistics to set tighter constraints on cosmological parameters. For this purpose, we have run N-body simulations of 5 models selected along the ($\sigma_8$-$\Omega_m$)-degeneracy and we have examined different non-Gaussian statistical tools in different representations in order to compare the discrimination power of each one. Using non-Gaussian statistics, we have looked for non-Gaussian signal present at small scales that is due to gravitational clustering.

The main conclusions of our analysis are the following:
\begin{enumerate}
\item The isotropic wavelet transform has been found to be the best representation of the non-Gaussian structures in weak lensing data allowing a better discrimination. 

\item We have shown that a wavelet transform denoising method like the MRLens filtering, which reconstructs essentially the non-Gaussian structures (the clusters), helps the statistics to better characterize the non-Gaussianity. 

\item We have introduced a new statistic called Wavelet Peak Counting (WPC) that consists in estimating a cluster count per scale of an isotropic wavelet transform.

\item $WPC$ has been found to be the best statistic compared to the others that we have tested (skewness, kurtosis, bispectrum, HC, $P_c$) and we have shown that this statistic estimated on MRLens filtered maps provide a strong discrimination between the 5 selected models.

\end{enumerate}

In this paper, we have compared systematically a wide range of non-Gaussian statistics proposed in the literature and $WPC$ have been found to be the best statistic. The comparison being done on models selected along the ($\sigma_8$-$\Omega_m$)-degeneracy, this study shows that the power spectrum (or the two-point correlation function) and WPC should therefore be used simultaneously. 
However, these two statistics probe the same field, therefore the correlations introduced by the combined measurement need to be taken into account \citep[see][]{correl:takada07}. This will be investigated in the future. 

Another issue, discussed in \S \ref{discussion}, is the selection effects. The selection function is needed for the estimation of the cosmological parameters and an accurate estimation of their errors. This study will be done in a future work.

Finally, while peak counting and WPC provide a lot of information, further statistics such as the cluster count per mass, the spatial cluster correlation will provide further constraints. Future work will be needed to fully exploit this approach.







\section*{Appendix A: The FDR method}
\label{annexea}

\PS{False Discovery Rate (FDR) is a statistical approach to the multiple testing problem, introduced by \cite{wlens:benjamini95}.

The FDR method offers an effective way to select an adaptative threshold, without any assumption. The FDR threshold is determined from the observed p-value distribution, and hence is adaptive to the amount of signal in the data.

This technique has been described by \cite{wlens:miller01,wlens:hopkins02,wlens:starck06,wlens:pires06} with several examples of astrophysical applications.
Instead of controlling the chance of any false positives, FDR controls the expected proportion of false positives. The FDR is given by the ratio (\ref{fdr}), that is, the proportion of declared
active which are false positives:
\begin{eqnarray} 
\mathcal{FDR} = \frac{V_{ia}}{D_a}
\label{fdr}
\end{eqnarray}
where $V_{ia}$ is the number of pixels truly inactive declared active,
and $D_a$ is the number of pixels declared active.
The FDR formalism ensures that, {\it on average}, the False Discovery Rate is no larger than
$\alpha$ which lies between 0 and 1. This procedure guarantees control over the FDR
in the sense that:
\begin{eqnarray} 
\mathcal{E(FDR)} \leq \frac{T_i}{V}.\alpha \leq \alpha
\end{eqnarray}

The unknown factor $\frac{T_i}{V}$ is the proportion of truly inactive pixels where $T_i$ is the number of inactive pixels and $V$ the total number of pixels.

The FDR procedure is as follows :\\
Let $P_1,..., P_n$ denote the p\_values from the N tests, listed from smallest 
to largest.\\ Let : 
\begin{eqnarray} 
d=max \big\{k:P_k < \frac{k.\alpha}{c_N.N}\big\}
\end{eqnarray}
where $c_N=1$, if p\_values are statistically independants.\\
Now, declare actived all the pixels with p\_values less than or equal 
to $P_d$.\\
Graphically, this procedure corresponds to plotting the $P_k$ versus 
$\frac{k}{N}$, superposing the line through the origin of slope 
$\frac{\alpha}{c_N}$ (see Fig. \ref{pltfdrl}), and finding the last point at which $P_k$ falls 
below the line, termed $P_d$. From this p\_value $P_d$, we can derive 
a threshold $\mathcal{T}$. All the pixels greater than $\mathcal{T}$ have 
a p\_value less than $P_d$ and are declared actives.

\begin{figure}[htp!]
\centerline{
\psfig{figure=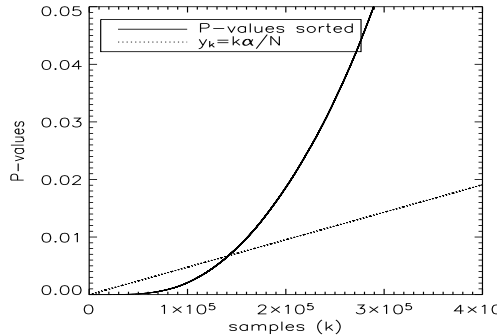,bbllx=3.cm,bblly=5.5cm,bburx=19.cm,bbury=23.cm,height=5.cm,width=6.5cm,clip=}
}
\caption{Finding a threshold graphically using the FDR procedure}
\label{pltfdrl}
\end{figure}

}

\section*{Appendix B: The Higher Criticism definition}
\label{annexeb}
 To define HC, first we convert the individual $\kappa_i$ into $p$-$values$.
Let $p_i = P\{N(0, 1) > \kappa_i\}$ be the $i$th $p$-$value$, and let $p_{(i)}$ denoted the $p$-$values$ sorted in increasing order. The higher criticism statistic is defined as:
\begin{eqnarray}
HC_n^*=\max_i\left|\frac{\sqrt n [i/n-p_{(i)}]}{\sqrt{p_{(i)}(1-p_{(i)})}}\right|.
\label{HC1}
\end{eqnarray}
Or in a modified  form :

\begin{eqnarray}
HC_n^+=\max_{i:1/n \leqslant p_{(i)} \leqslant 1-1/n}\left|\frac{\sqrt n [i/n-p_{(i)}]}{\sqrt{p_{(i)}(1-p_{(i)})}}\right|.
\label{HC2}
\end{eqnarray}

\section*{Appendix C: Description of the representations}
\label{annexec}
\subsection*{The anisotropic bi-orthogonal wavelet transform}
The most commonly used wavelet transform is the undecimated bi-orthogonal wavelet transform (OWT).
Using the OWT, a 2D signal S can be decomposed as follows :
\begin{eqnarray}
S(x,y) = \sum_{k_x, k_y}  \phi_{J, k_x, k_y}(x,y) C_J(k_x, k_y)  \\
+ \sum_d \sum_{k_x, k_y} \sum_{j=1}^{J} \psi^d_{j,k_x,k_y} (x, y) w^d_{j}(k_x, k_y),
\label{otw} 
\end{eqnarray}
with where $\phi$ and $\psi^d$ are respectively the scaling function and the wavelet functions that prioritize the horizontal, vertical and diagonal directions. $J$ is the number of resolutions used in the decomposition, $w^d_j$ the wavelet (or detail) coefficients at scale $j$ and direction $d$, and $C_J$ is a smooth version of the original signal $S$. 

\subsection*{The undecimated isotropic wavelet transform}
The Undecimated Isotropic Wavelet Transform (UIWT) decomposes an $n \times n$ image $\kappa$ as in :
\[ 
\kappa(x,y)= C_J (x,y) + \sum_{j=1}^{J} w_j (x,y), 
\]
where $C_{J}$ is a coarse or smooth version of the original image $\kappa$ and $w_j$ represents the details in $\kappa$ at scale $2^{j}$ \citep[see][for details]{starck:book06}. 

\subsection*{The ridgelet transform}
The classical multiresolution ideas only address a portion of the
whole range of interesting phenomena: the roughly isotropic one at all
scales and all locations. The ridgelet transform have been proposed 
as an alternative to the wavelet representation of image data.

Given a function $f(x_1,x_2)$, the ridgelet transform is the superposition
of elements of the form $a^{-1/2}\psi((x_1\cos\theta+\sin\theta-b)/a)$, $\psi$ is the wavelet,
$a>0$ the scale parameter, $b$ the location parameter and $\theta$ the
orientation parameter. The ridgelet is constant along lines
$x_1\cos\theta+x_2\sin\theta=const$, and transverse to these ridges it
is a wavelet. 

\subsection*{The curvelet transform}
Ridgelets are essentially focused on dealing with straight lines rather than curves, ridgelets can be adapted to representing objects with curved edges using an appropriate multiscale
localization. If one uses a sufficiently fine scale to capture curved edges, such edges 
are almost straight. As a consequence the curvelet transform has been 
introduced, in which ridgelet are used in a localized manner.

The idea of the curvelet transform \citep{cur:candes99,cur:starck03} is to first decompose the image into
a set of wavelet planes, then to decompose each plane in several blocks
(the block size can change at each scale level) and to analyse each block 
with a ridgelet transform.
The finest the scale is, the more sensitive to the curvature the analysis is.
As a consequence, curved singularities can be well approximated with very 
few coefficients.

\section*{Appendix D: The MRLens filtering}
\label{annexed}
\PS{The MRLens filtering \citep{wlens:starck06} is a non linear filtering based on the Bayesian theory that searches for a solution that maximizes the a posteriori probability.
Choosing the prior is one of the most critical aspects of the Bayesian analysis. 
The MRLens filtering uses a multiscale entropy prior.

Assuming Gaussian noise, the MRlens filtering solves the following minimization :
\begin{eqnarray}
J(\kappa)= \frac{\parallel \kappa_n   - \kappa \parallel ^2}
  {2\sigma_n^2} 
  + \beta \sum_{j=1}^{J} \sum_{k,l} h_n( ({\cal W} \kappa)_{j,k,l})    
\end{eqnarray}
where $\sigma_n$ the noise standard deviation, $J$ the number
of scales,  $\beta$ is the regularization parameter and $\cal W$ is the Wavelet Transform operator.

Full details of the minimization 
algorithm can be found in \cite{starck:sta01_1}, as well as the way to determine automatically
the regularization parameter $\beta$.  

In \cite{wlens:starck06}, it has been shown that the MRLens filtering outperforms the existing methods.
The MRLens filtering has already been used for several applications on weak lensing data and especially, it has been selected to filter the dark matter mass map obtained by the Hubble Space Telescope in the COSMOS field.
The complete MRLens software package to perform weak lensing filtering can be downloaded from http://www-irfu.cea.fr/Ast/878.html. 
}

\bibliographystyle{astron}
\bibliography{discrimination_revised}

\end{document}